\documentclass[]{pasj02} 
\usepackage[switch,mathlines]{lineno} 
\usepackage{caption}
\usepackage{graphicx}
\usepackage{subcaption}
\usepackage{natbib} 
\usepackage{xcolor}%

\jyear{2025}
\Received{}
\Accepted{}

\begin{document} 

\title{
XRISM mock observations of simulated AGN jets in the core of a galaxy cluster
}

\author{
 Mahiro \textsc{Shirotori},\altaffilmark{1} \email{shirotori-mahiro@ed.tmu.ac.jp} 
 Yutaka \textsc{Fujita}\altaffilmark{1}
}
\altaffiltext{1}{Department of Physics, Graduate School of Science
Tokyo Metropolitan University
1-1 Minami-Osawa, Hachioji-shi, Tokyo 192-0397}




\maketitle

\begin{abstract}
Jets from active galactic nuclei (AGNs) are expected to heat the surrounding intracluster medium (ICM). We investigate how the interaction between jets and the ICM appears in high-resolution X-ray observations using mock X-ray observations based on two-dimensional hydrodynamic simulations.
We constructed a model of an active galactic nucleus (AGN) similar to Cygnus A (Cyg A), a powerful FR II radio galaxy. Our simulations model bipolar jets propagating into a stratified ICM, forming forward shocks and low-density cocoons. Based on these results, we generate synthetic spectra that incorporate both shocked and unshocked ICM components. Then, we perform mock observations using the XRISM/Resolve X-ray spectrometer.  We focus particularly on viewing angle effects.
Our mock observations revealed that the smallest line broadening, observed as velocity dispersion, associated with the cocoon's bulk expansion occurs when observing along the jet direction, where the expansion velocity is highest. Although this may appear counterintuitive, it occurs because the rapidly expanding jet head contributes little to X-ray emission due to its high temperature and low density.
Our results highlight the importance of considering the temperature and density structure of AGN-driven shocks and cocoons when interpreting XRISM data. These findings lay the groundwork for XRISM's observations of AGN jets and will improve our understanding of AGN feedback processes in galaxy clusters.
\end{abstract}


\section{Introduction}

Clusters of galaxies are the largest gravitationally bound structures in the universe, consisting of hundreds to thousands of galaxies, dark matter, and the hot, diffuse plasma known as the ICM \citep{Sarazin1986}. The ICM typically has a temperature of $10^7-10^8~\mathrm{K}$ and emits strongly in X-rays through thermal bremsstrahlung and line emission. The cooling time in cluster cores is short, on the order of $10^8-10^9~\mathrm{yr}$. If radiative cooling is not prevented, a large amount of star formation and molecular gas accumulation would be expected in central cluster galaxies \citep{Cowie1977, Fabian1977}. However, such massive reservoirs of cooled gas have not been observed (e.g., \citealt{Peterson2001, Peterson2003, Tamura2001, Kaastra2001, Lewis2002}), giving rise to the so-called "cooling flow problem". A leading solution is feedback from the central active galactic nucleus (AGN; \citealt{Fabian2012, McNamara2007}).

AGNs release enormous energy through accretion onto supermassive black holes, with relativistic jets being one of the most important channels. These jets penetrate the ICM and generate large-scale structures such as cocoons. X-ray observations have revealed that they leave cavities, excite weak shocks, and drive ripples in the ICM (e.g., Perseus: \citealt{Boehringer1993, Fabian2003}; Hydra A: \citealt{Nulsen2005}; M87: \citealt{Forman2007}). Such processes not only offset radiative cooling but also influence the thermodynamic and structural evolution of clusters \citep{Birzan2004, Dunn2006, Rafferty2006, McNamara2007}.

Theoretical studies of jet--ICM interaction have been developed through both analytical models and numerical simulations. Early theoretical frameworks described the evolution of expanding cocoons and radio lobes inflated by AGN jets \citep{Scheuer1974, Begelman1989, Kaiser1997}, which provided quantitative predictions of cavity sizes and expansion velocities. More recently, numerical simulations have examined the excitation of shocks, the development of turbulence, and the growth of cocoon structures (e.g., \citealt{ONeill2010, Gaspari2012, Mendygral2012}). Furthermore, cosmological simulations have been used to reproduce Perseus-like clusters and to perform mock X-ray observations, bridging theory and observation in the study of AGN feedback \citep{Truong2024}.

Cyg A is a particularly important target in the study of AGN feedback. It is a prototypical Fanaroff-Riley (FR) II radio galaxy \citep{Fanaroff1974, Carilli1996}, hosting powerful jets with kinetic powers about $10^{46}~\mathrm{erg\,s^{-1}}$ \citep{Wilson2006, Ito2008, Snios2018}. These jets have been observed to produce prominent X-ray cavities and shock fronts \citep{Smith2002, Wilson2006, Snios2018}. Owing to its proximity ($z=0.056075$; \citealt{Owen1997}) and high brightness, Cyg A provides an ideal laboratory for detailed studies of AGN jet--ICM interactions.

Direct measurements of ICM gas motions driven by AGNs were long out of reach. A breakthrough came with the Hitomi observation of the Perseus cluster, where the Soft X-ray Spectrometer (SXS) directly measured a velocity dispersion of $\sim 160~\mathrm{km\,s^{-1}}$, demonstrating that AGN-driven gas motions can be relatively gentle \citep{Hitomi2016}. A subsequent analysis with spatially resolved spectroscopy further revealed variations in the velocity structure across the cluster core \citep{Hitomi2018}, establishing a more detailed picture of gas dynamics in cool-core systems.

The successor mission, XRISM \citep{Tashiro2021}, is equipped with the high-resolution Resolve spectrometer \citep{Ishisaki2022}, which provides $\sim5~\mathrm{eV}$ energy resolution and enables precise measurements of ICM dynamics. Recent XRISM observations have begun to reveal the dynamical properties of ICM affected by AGN feedback. For example, in Hydra A, despite hosting jets nearly an order of magnitude more powerful than those in Perseus, the velocity dispersion within the cluster core was found to be only $164 \pm 10~\mathrm{km\,s^{-1}}$, similar to the Hitomi Perseus measurement, suggesting that turbulent dissipation alone may be insufficient to counteract cooling across the entire cooling volume \citep{Rose2025}.
In Perseus, radial mapping revealed that the velocity dispersion decreases to $\sim 70~\mathrm{km\,s^{-1}}$ at $\sim 70$ kpc, compared to $\sim 170~\mathrm{km\,s^{-1}}$ within 60 kpc, before rising again to $\sim 190~\mathrm{km\,s^{-1}}$ in the outer regions, with the dip likely associated with a cold front \citep{XRISM-Perseus2025}.

In this study, we perform two-dimensional hydrodynamic simulations of jets modeled after Cyg A, and construct mock XRISM/Resolve spectra. We specifically examine how bulk expansion depends on jet orientation relative to the line-of-sight (LOS). Our goal is to evaluate to what extent bulk expansion can be detected with high-resolution X-ray spectroscopy. For this purpose, we adopt simplified simulations that focus on the detectability of bulk motions. This work provides theoretical guidance for interpreting future XRISM observations of Cyg A and contributes to a deeper understanding of AGN feedback mechanisms in galaxy clusters.

In this paper, we adopt a $\Lambda$CDM cosmology with $\Omega_m = 0.315$ and $H_0 = 67.4~\mathrm{km~s^{-1}~Mpc^{-1}}$. At a redshift of $z = 0.056075$, corresponding to Cyg A, the angular scale is $1.13~\mathrm{kpc~arcsec^{-1}}$.

\section{Simulation setups}
\subsection{Jets}

We perform two-dimensional, axisymmetric hydrodynamic simulations of jet propagation using the CANS+ magnetohydrodynamic (MHD) code \citep{Matsumoto2019} though we ignore magnetic fields for simplicity. The computational domain spans $0~ \leq x \leq 40~\mathrm{kpc}$ and $0 \leq z \leq 90~\mathrm{kpc}$. The $z$-axis is the axis of symmetry. $0.2~\mathrm{kpc} \times 0.2~\mathrm{kpc}$, with $300 \times 450$ cells. A supersonic jet is injected in the $+z$ direction from the origin. To model the jet's bipolar nature, boundary conditions are imposed such that $v_z$ is reflected across the $x$-axis and $v_x$ is reflected across the $z$-axis. Neumann boundary conditions (zero gradient) are used elsewhere.

Assuming negligible internal energy, only kinetic energy is injected. The power of the one-sided jet, $L_{\mathrm{jet}}$, is expressed as follows:
\begin{equation}
L_{\mathrm{jet}} = \frac{\pi}{2} \rho_{\mathrm{jet}} v_{\mathrm{jet}}^3 r_{\mathrm{jet}}^2\label{jetpower},
\end{equation}
where $\rho_{\mathrm{jet}}$, $v_{\mathrm{jet}}$, and $r_{\mathrm{jet}}$ are the density, velocity, and radius of the jet, respectively.

 Based on observations of Cyg A, we adopted $L_{\mathrm{jet}} = 10^{46}~\mathrm{erg~s^{-1}}$. VLA observations suggest a jet radius of $r_{\mathrm{jet}} = 0.1-1~\mathrm{kpc}$ around 1 kpc from the nucleus \citep{Nakahara2019}. Therefore, we adopt $r_{\mathrm{jet}} = 1~\mathrm{kpc}$.
Following \citet{Ohmura2023a}, we set the velocity of the jet to $v_{\mathrm{jet}} = 0.3c$, where $c$ is the speed of light. The jet density $\rho_{\mathrm{jet}}$ is calculated from equation (\ref{jetpower}).
The initial conditions and jet parameters adopted in the simulation are summarized in Table~\ref{tab:table1}.

\begin{table}
  \caption{Initial conditions and jet parameters used in the simulation.}
  \label{tab:table1}
  \centering
  \begin{tabular}{lc}
    \hline
    Parameter & Value \\
    \hline
    $\rho_0$ & $1.36 \times 10^{-25}~\mathrm{g~cm^{-3}}$ \\
    $r_c$ & $20~\mathrm{kpc}$ \\
    $\beta$ & $0.5$ \\
    $kT$ & $6~\mathrm{keV}$ \\
    $r_{\mathrm{jet}}$ & $1~\mathrm{kpc}$ \\
    $v_{\mathrm{jet}}$ & $0.3c$ \\
    $L_{\mathrm{jet}}$ & $ 10^{46}~\mathrm{erg~s^{-1}}$ \\
    \hline
  \end{tabular}
\end{table}

\subsection{ICM}
\label{sec:ICM}

We assume that the unshocked ICM is spherically distributed and in hydrostatic equilibrium.
The ICM's mass density profile is described by the $\beta$-model \citep{Cavaliere1976}:
\begin{equation}
\rho(r) = \rho_0 \left[1 + \left( \frac{r}{r_{\mathrm{c}}} \right)^2 \right]^{-3\beta/2},
\label{density}
\end{equation}
where $r = \sqrt{x^2 + y^2 + z^2}$, and $\rho_0$, $r_{\mathrm{c}}$, and $\beta$
denote the core radius, the central density, and the so-called $\beta$ parameter, respectively.
Following Chandra observations of Cyg~A \citep{Snios2018}, we set $\rho_0 = 1.36 \times 10^{-25}~\mathrm{g~cm^{-3}}$, $r_{\mathrm{c}} = 20~\mathrm{kpc}$, and $\beta = 0.5$.

Assuming a uniform ICM temperature of $T = 6~\mathrm{keV}$, based on Chandra measurements \citep{Smith2002}, the pressure profile is given by:
\begin{equation}
P(r) = \frac{kT}{\mu m_{\mathrm{p}}} \rho(r)\label{EOS},
\end{equation}
where $k$ is the Boltzmann constant, $\mu=0.6$ is the mean molecular weight, and $m_{\mathrm{p}}$ is the proton mass.
The gravitational acceleration under hydrostatic equilibrium is:
\begin{equation}
g(r) = \frac{1}{\rho} \frac{dP}{dr}.
\end{equation}
From equation (\ref{EOS}), the gravitational acceleration can be rewritten as follows:
\begin{equation}
g(r) = -\frac{kT}{\mu m_{\mathrm{p}}} \frac{3\beta r}{r_{\mathrm{c}}^2 \left[1 + \left(\frac{r}{r_{\mathrm{c}}}\right)^2\right]}.
\end{equation}

Recent X-ray observations have revealed the existence of low-level turbulence in the ICM ($\lesssim 200\rm\: km\: s^{-1}$; e.g., \citealt{Hitomi2016, XRISM-Perseus2025, Rose2025}). However, for simplicity's sake, we do not include it in the background (unshocked) ICM in our simulations. Since the shock front propagates much faster in our simulations ($\gtrsim 3500\rm\: km \:s^{-1}$), the turbulence would not affect the results.

\section{Simulation Results}

\begin{figure*}
  \centering
  \begin{subfigure}[b]{0.24\textwidth}
    \centering
    \includegraphics[height=6.7cm]{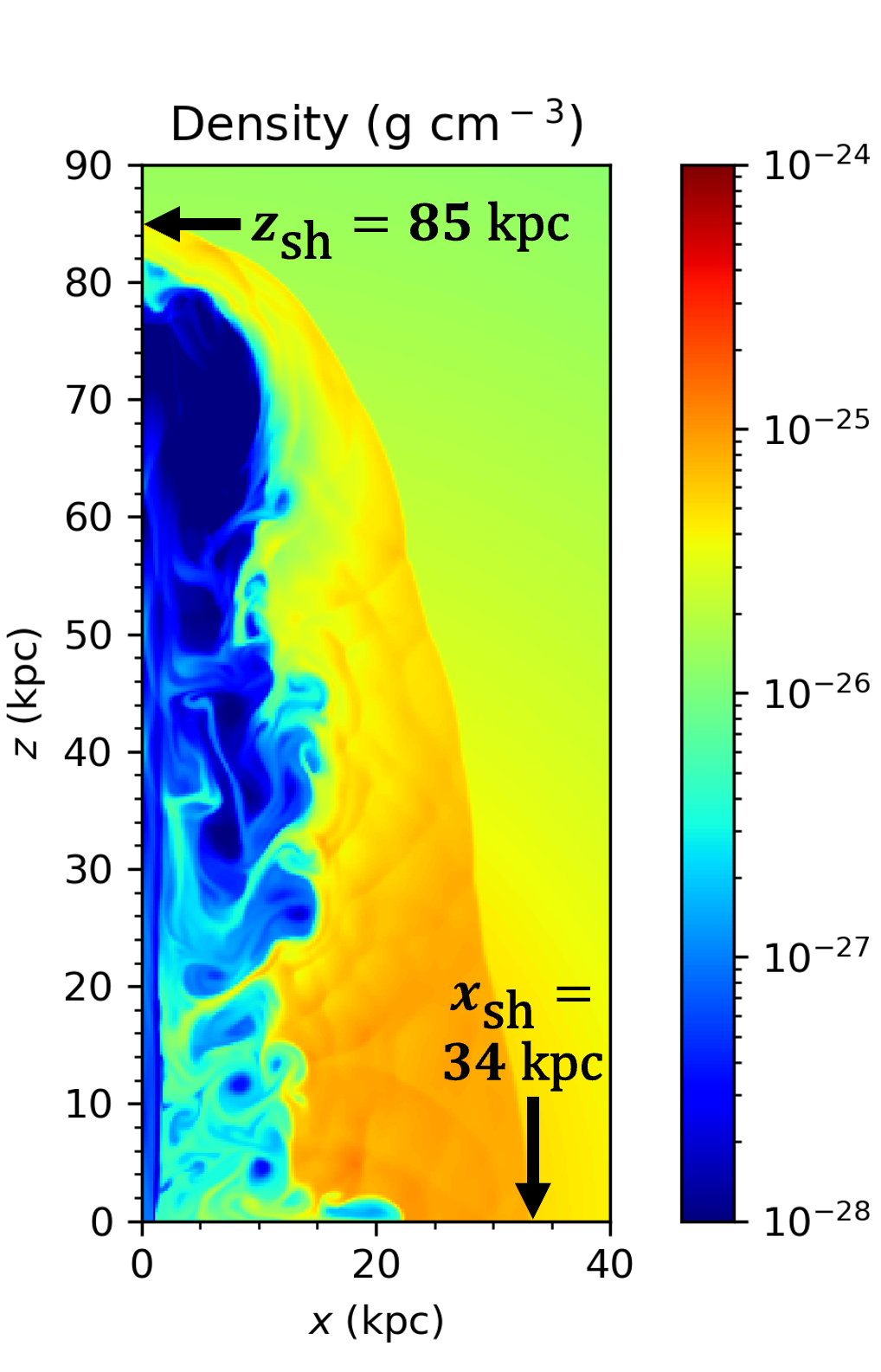}
    \subcaption{}
    \label{fig:fig1a}
  \end{subfigure}
  \hspace{1em}
  \begin{subfigure}[b]{0.24\textwidth}
    \centering
    \includegraphics[height=6.7cm]{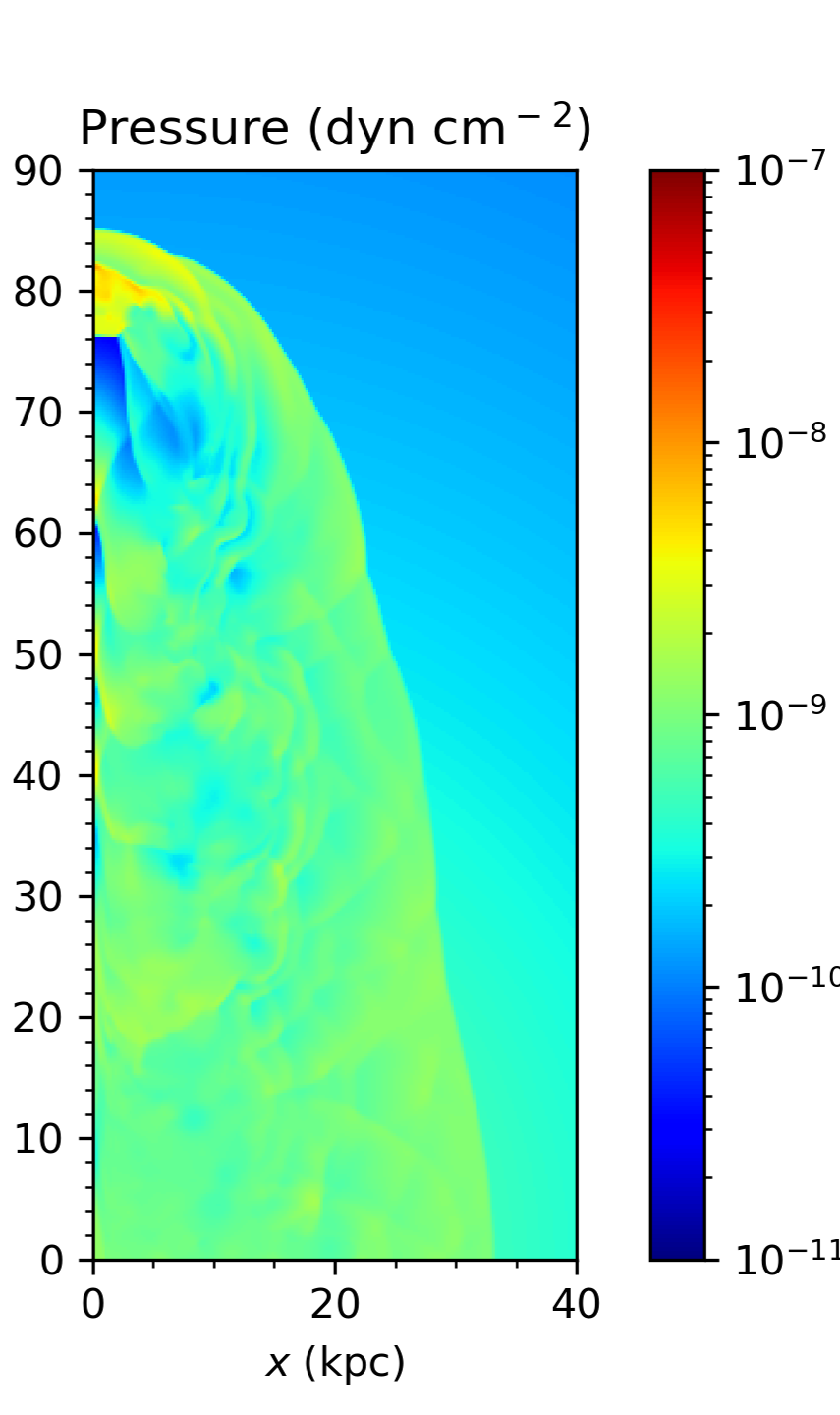}
    \subcaption{}
    \label{fig:fig1b}
  \end{subfigure}
  \begin{subfigure}[b]{0.24\textwidth}
    \centering
    \includegraphics[height=6.7cm]{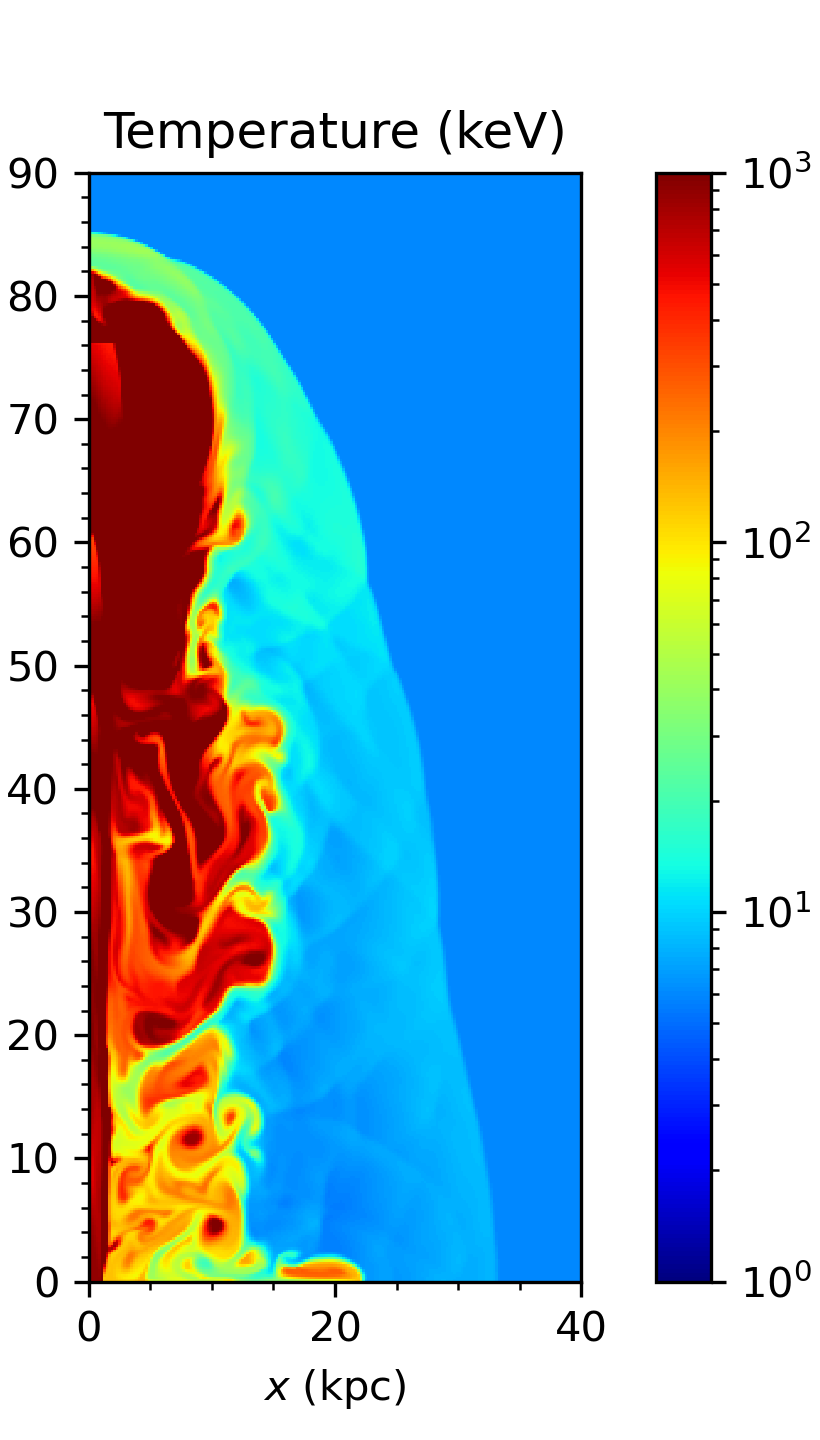}
    \subcaption{}
    \label{fig:fig1c}
  \end{subfigure}
  \begin{subfigure}[b]{0.24\textwidth}
    \centering
    \includegraphics[height=6.7cm]{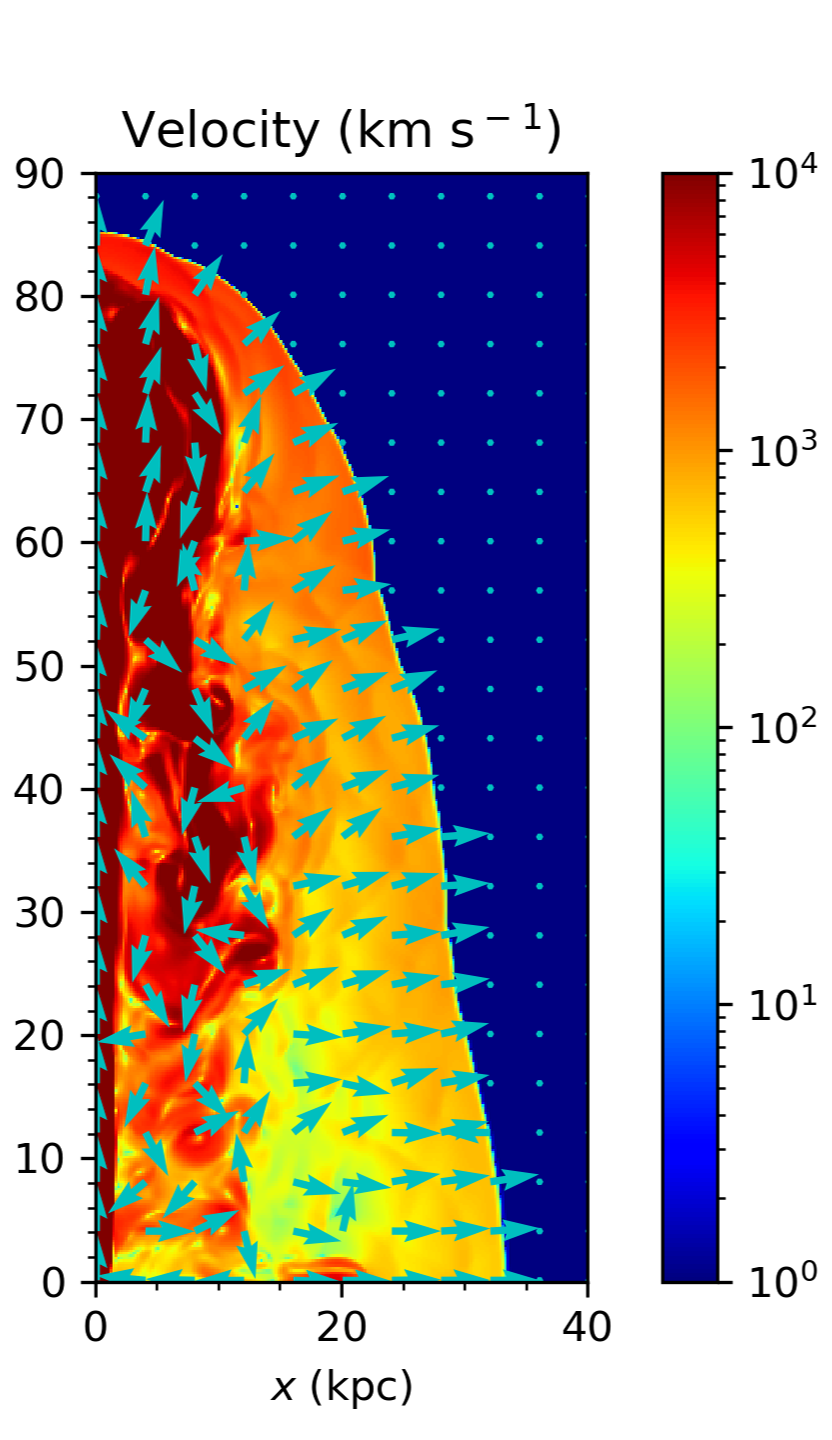}
    \subcaption{}
    \label{fig:fig1d}
  \end{subfigure}
  \caption{
  Simulation results at $t = 13.5~\mathrm{Myr}$. From left to right: density, pressure, temperature, and velocity maps.
  The velocity map shows the magnitude of the velocity vector in color and its direction with arrows. The assumed axisymmetry yields $v_y = 0~\mathrm{km~s^{-1}}$. The semi-minor and semi-major axes of the forward shock are $x_{\mathrm{sh}} = 34~\mathrm{kpc}$ and $z_{\mathrm{sh}} = 85~\mathrm{kpc}$, respectively.
  {Alt text: Simulation results in the computational domain.}
  }
  \label{fig:fig1}
\end{figure*}

\subsection{Jet age and size}

We adjusted the age of the simulated jet to make its size comparable to that of Cyg A. Regardless of the viewing angle, the apparent and physical semi-minor axes of the forward shock or cocoon formed around the jet are nearly identical. Thus, in figure~\ref{fig:fig1}, we present density, pressure, temperature, and velocity maps at $t = 13.5$~Myr, when the semi-minor axis of the simulated forward shock, $x_{\mathrm{sh}}$, is comparable to the observed value, $x_{\mathrm{obs}}\sim 34$~kpc for Cyg A \citep{Snios2018}. 
The jet injection time derived in this work ($13.5$ Myr) lies within the range of previous theoretical estimates from cocoon dynamics models ($3\!-\!30$ Myr; \citealt{Kino2005}) and is comparable to the observational estimate from \textit{Chandra} data ($\sim 20$ Myr; \citealt{Snios2018}). Our result is therefore broadly consistent with earlier studies, although differences in the definition of source age and methodological assumptions should be kept in mind.
In contrast, the apparent semi-major axis depends on the viewing angle. VLBI observations suggest that the viewing angle of Cyg A lies between $\theta=35^\circ$ and $80^\circ$, where $\theta=0$ is the jet direction \citep{Bartel1995}. 
\citet{Ohmura2023b} estimated the viewing angle using two-temperature MHD simulations and mock X-ray observations. By comparing the simulated and observed X-ray surface brightness profiles and temperature jumps, they determined that the angle between the jet axis and the LOS is between $\theta=35^\circ$ and $55^\circ$.
We assume that the fiducial value of the viewing angle is $\theta = 53^\circ$ in the mock observations in section~\ref{sec:mock}. Since the apparent semi-major axis is $z_{\rm obs}\sim 68$~kpc \citep{Snios2018}, the physical value should be $z_{\rm sh}\sim z_{\rm obs}/\sin{\theta}\sim 85$~kpc, consistent with the simulation results in figure~\ref{fig:fig1}.

\subsection{Jet and shocked gas}

The region along the $z$-axis (blue region in figure~\ref{fig:fig1a}) exhibits a turbulent structure created by the collision between the jet and the surrounding ICM. A forward shock is generated ahead of the jet where the matter is scattered. This turbulent region is extremely hot ($>200$~keV; figure~\ref{fig:fig1c}) and cannot be detected by X-ray satellites such as XRISM. Instead, it should appear as a cavity.

Outside the turbulent region, the region behind the forward shock can be divided into two regions: a lower-density region at $x\gtrsim 15$~kpc and $z \gtrsim 40$~kpc (yellow region in figure~\ref{fig:fig1a}) and a higher-density region at $x\gtrsim 15$~kpc and $z \lesssim 40$~kpc (orange region in figure~\ref{fig:fig1a}). The velocity map (figure~\ref{fig:fig1d}) shows that the shocked ICM is expanding outward. 
This expansion is governed by the high internal pressure of the cocoon, which is consistent with analytical model predictions \citep{Begelman1989}.

In the lower-density region, the gas moves outward at a significantly higher velocity than in the higher-density region. In particular, the region near the jet tip moves almost along the $z$-axis at velocities up to $\sim 3500~\mathrm{km~s^{-1}}$. The temperature map (figure~\ref{fig:fig1c}) shows that this region is hotter than the higher-density region, reaching temperatures of several tens of keV. This heating is caused by the conversion of the jet's kinetic energy into thermal energy via strong shocks at the jet head that compress and heat the ambient ICM \citep{Kaiser1997}. 
Conversely, the higher-density region expands more slowly at a few hundred $\mathrm{km~s^{-1}}$, and the velocity component in the $x$-direction is greater than in the $z$-direction. This region has a lower temperature than the low-density region, but it is still higher than that of the unshocked ICM.

\section{Mock Observation}
\label{sec:mock}
\subsection{Generation of Spectra}\label{ssec:31}

\begin{figure}
\centering
\captionsetup{justification=raggedright}
\includegraphics[width=0.85\linewidth]{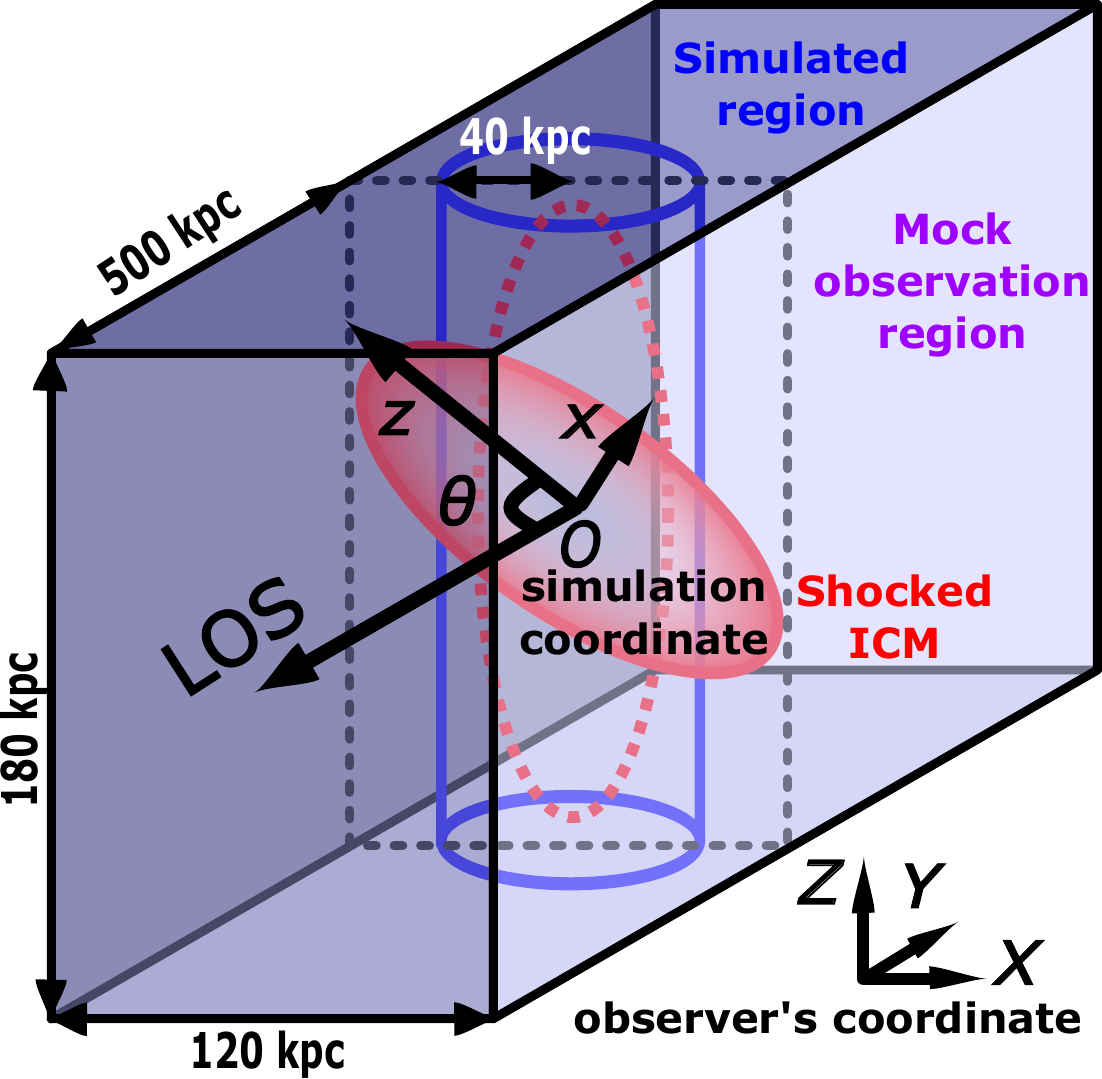}
\caption{
Schematic illustration of the mock observation setup. The outermost rectangular prism indicates the full observation volume, which has an observation area of $120~\mathrm{kpc} \times 180~\mathrm{kpc}$ on the $XZ$ plane.
The $Y$-axis is in the direction of the LOS. The blue cylinder shows the simulated region when the jet direction ($z$-axis) is parallel to the $Z$-axis. The viewing angle is defined by the angle between the $z$-axis and $Y$-axis. The red dashed ellipse represents the shocked ICM (cocoon) at $\theta = 90^\circ$, while the red  solid ellipse corresponds to the shocked ICM at $\theta \neq 90^\circ$.
{Alt text: Schematic illustration. The cylinder has a height of 180 kpc, with its top and bottom faces coincident with the top and bottom $XY$ planes of the rectangular prism. The cylinder’s central axis passes through the origin $O$.}
}
\label{fig:fig2}
\end{figure}

Figure~\ref{fig:fig2} shows a schematic of the mock observation setup. We introduce the observer's coordinate system ($XYZ$), which differs from the simulation coordinate system ($xz$). The $Y$-axis is in the direction of the LOS. Note that our simulations are axis-symmetric around the $z$-axis. 
In addition to the emission from the simulation region, we consider the emission from the unshocked ICM outside the simulation region. The observation area spans $\sim 120~\mathrm{kpc} \times 180~\mathrm{kpc}$ in the $XZ$ plane, corresponding to the $\sim 4 \times 6$ pixel field of view of XRISM/Resolve at the distance of Cyg~A. 
We set the depth in the $Y$ direction ($\pm500$ kpc) to be the same as the radius ($\sim500$ kpc) within which X-rays from Cyg A are observed \citep{Smith2002}. The origin $O$ of the observer coordinate system is set to coincide with that of the simulation coordinate system. The blue cylinder indicates the simulated region when the $z$- and $Z$-axes are parallel. This is obtained by mirroring the simulation domain along the $x$-axis and rotating it about the $z$-axis. The viewing angle, $\theta$, is defined by the angle between the $z$-axis (jet axis) and the $Y$-axis (LOS). When $\theta \neq 90^\circ$, the shocked ICM region or cocoon tilts against the $Z$-axis (red solid ellipse in figure~\ref{fig:fig2}).

We generate mock XRISM spectra by dividing the region of interest into grids and synthesizing a spectrum from each grid cell. We assume that the X-ray emission arises from optically thin thermal plasma.  For the shocked ICM inside the forward shock, we used the \texttt{bapec} spectral model in XSPEC v12.13.1 \citep{Arnaud1996}, which accounts for the effects of turbulence as line broadening. Cells with $\sqrt{v_x^2 + v_z^2} > 1~\mathrm{km~s^{-1}}$ are classified as shocked ICM. 

To reduce computation time, the $5 \times 5$ simulation grids are merged into one $1~\mathrm{kpc} \times 1~\mathrm{kpc}$ block. The averaged physical quantities, denoted by the symbol $q$, are then calculated for each block.
\begin{equation}
\langle q \rangle = \frac{\int_{\text {block}} w\,q(x,z)\,dV}{\int_{\text {block}} w\,dV},
\end{equation}
where $w = \rho^2 T^{1/2}$ represents the emissivity of thermal bremsstrahlung. The results are not sensitve to the details of the weighing factor $w$.  The integration is performed for each block. If a block contains both shocked and unshocked components, the integration is performed separately for each.
For the shocked ICM, each $1~\mathrm{kpc} \times 1~\mathrm{kpc}$ block forms a ring around $z$-axis in the three-dimensional space. The ring is divided into 36 azimuthal sectors at $10^\circ$ intervals to compute LOS velocities. The LOS velocity and the averaged LOS velocity for sector $i$ is given by:
\begin{equation}
v_{\mathrm{LOS}} = v_x \sin \theta \cos\left(\frac{2\pi i}{36}\right) + v_z \cos \theta  ,
\end{equation}
\begin{equation}
\langle v_{\mathrm{LOS}} \rangle = \langle v_x \rangle \sin \theta \cos\left(\frac{2\pi i}{36}\right) + \langle v_z \rangle \cos \theta  ,
\end{equation}
where $i = 0, 1, \dots, 35$.
We considered the turbulent gas motion within each grid cell and calculated the gas velocity dispersion for each block as follows:
\begin{equation}
\sigma_\mathrm{block}^{2} = \frac{\int_{\text {block}}\left(v_{\text {LOS}}-\langle v_{\mathrm{LOS}} \rangle\right)^{2} w d V}{\int_{\text {block}} w d V}.
\end{equation}

Since the unshocked ICM outside the forward shock is static, we manually generate its spectra without using simulation results. For the unshocked ICM, the density profile follows equation~(\ref{density}), and the temperature profile is approximated from that in \citet{Snios2018}:
\begin{equation}
T(r) =
\begin{cases}
6~\mathrm{keV}, & r \leq 100~\mathrm{kpc}, \\
8~\mathrm{keV}, & r > 100~\mathrm{kpc}.
\end{cases}
\end{equation}
Note that the simulated region lies within $r < 100~\mathrm{kpc}$, so the initial temperature is $6~\mathrm{keV}$.
Our simulations did not include low-level turbulence in the unshocked ICM (section~\ref{sec:ICM}), but this turbulence could affect the X-ray spectra when the LOS contribution of the unshocked ICM is more significant than that of the shocked ICM. Thus, for the unshocked ICM, we use the \texttt{bapec} spectral model. 
We add turbulence to the mock spectra with a LOS velocity dispersion of $\sigma_0=160~\mathrm{km~s^{-1}}$, based on Hitomi observations of the Perseus cluster \citep{Hitomi2016}.

Mock spectra are generated for each block of the shocked ICM and for the entire unshocked ICM using the \texttt{fakeit} command in XSPEC. 
This command simulates spectra based on plasma parameters and instrument responses. We adopt an exposure time of $300~\mathrm{ks}$ and use the XRISM/Resolve response files \texttt{rsl\_Hp\_5eV.rmf} and \texttt{rsl\_pointsourse\_GVclosed.arf}\footnote{https://xrism.isas.jaxa.jp/research/proposer/obsplan/response/index.html}. The generated spectra are co-added using \texttt{mathpha}. The ICM's metal abundance is fixed at $0.5~Z_\odot$ \citep{Snios2018}.
Since \texttt{bapec} models are limited to the temperature range of $0.008-64~\mathrm{keV}$, blocks with temperatures outside this range were excluded from the analysis. In addition to the combined spectra of the shocked and unshocked ICMs, we discuss the spectra of the shocked ICM alone for comparison (see section~\ref{sec:shocked}).

\subsection{Direct Estimation}
\label{sec:direct}

Physical parameters can be estimated directly from the simulation results, which can then be compared with mock observation results. The spectral temperature $T_{\mathrm{spec}}$ is estimated using the formula proposed in \citet{Mazzotta2004}, which is often used to compare simulation results with observations:
\begin{equation}
T_{\mathrm{spec}} = 
\frac{\int \rho^2 T^{-3/4} TdV}{\int \rho^2 T^{-3/4}dV}.
\end{equation}
The integration is performed on the total observation volume. This formula accurately reproduces the inferred temperatures of cluster gas above $3~\mathrm{keV}$ and solves the problem of overestimation caused by brightness-weighted temperatures.
The normalization parameter, which is compared to the "norm" parameter of the \texttt{bapec} model in mock observations, is calculated as follows:
\begin{equation}
\mathrm{norm} = 
\frac{10^{-14}}{4\pi \left[ D_{\mathrm{A}} (1+z) \right]^2}
\int n_{\mathrm{e}} n_{\mathrm{H}} dV,
\end{equation}
where $D_{\mathrm{A}}$ is the angular diameter distance, $z$ is the redshift, and $n_{\mathrm{e}}$, $n_{\mathrm{H}}$ are the electron and hydrogen number densities.

Two weighting schemes are used to calculate the velocity dispersion:

\begin{enumerate}
  \item Brightness-weighted based on thermal bremsstrahlung:
  \begin{equation}
  w = w_{\mathrm{ew}} = \rho^2 T^{1/2}
  \end{equation}

  \item Weighted by the population fraction of He-like Fe ions:

 At temperatures of several keV, He-like Fe ions dominate the emissivity of the relevant spectral lines and are the main determinant of gas velocity in spectral fitting. Thus, we use their temperature-dependent population fraction $f(T)$ to compute a velocity dispersion:
  \begin{equation}
  w = w_{\mathrm{Fe}} = \rho f(T)
  \end{equation}
  The function $f(T)$ is approximated based on the population fraction presented in the ASTRO-H white paper \citep[Figure 9a]{Smith2014} as:
  \begin{equation}
  f(T) =
  \begin{cases}
  0.76 \left( \frac{T}{4~\mathrm{keV}} \right)^{-0.71}, & 4~\mathrm{keV} \leq T < 6~\mathrm{keV}, \\
  0.57 \left( \frac{T}{6~\mathrm{keV}} \right)^{-1.61}, & 6~\mathrm{keV} \leq T < 10~\mathrm{keV}, \\
  0.25 \left( \frac{T}{10~\mathrm{keV}} \right)^{-2.35}, & 10~\mathrm{keV} \leq T \leq 64~\mathrm{keV}.
  \end{cases}
  \end{equation}
\end{enumerate}

The velocity dispersion for the shocked ICM alone is estimated as:
\begin{equation}
\label{eq:sig_sh}
\sigma_\mathrm{bulk,a}^{2} = \frac{\int_{\text {shocked}}\left(v_{\text {LOS}}-\bar{v}\right)^{2} w_\mathrm{a} d V}{\int_{\text {shocked}} w_\mathrm{a} d V}\:,
\end{equation}
and the velocity dispersion for the total ICM (shocked + unshocked ICM) is estimated as:
\begin{equation}
\label{eq:sig_tot}
\begin{aligned}   
&\sigma_\mathrm{total,a}^{2} =\\ 
&\frac{\int_{\text {shocked}}\left(v_{\text {LOS}}-\bar{v}\right)^{2} w_\mathrm{a} d V+\sigma_0^{2} \int_{\text {unshocked}} w_\mathrm{a} d V}{\int_{\text {shocked}+ \text {unshocked}} w_\mathrm{a} d V},
\end{aligned}
\end{equation}
where $w_{\mathrm{a}}$ is the weighting factor (a is either "ew" or "Fe"), $v_{\mathrm{LOS}}$ is the LOS velocity, and $\bar{v}$ is the average bulk velocity. Due to the symmetry of our simulations, we have $\bar{v}=0$. Note that XRISM's angular resolution is modest, with a relatively large point spread function of about $1.3'$ for the half-power diameter. Thus, Cyg A cannot be well resolved and is observed in one field in our setup (figure~\ref{fig:fig2}). 
Therefore, the velocity dispersion of the shocked ICM primarily originates from the bulk expansion of the cocoon rather than from small-scale turbulence.

\section{Discussion}
\label{sec:discuss}
\subsection{Shocked ICM  Mock Spectra}
\label{sec:shocked}

\begin{figure*}
  \begin{center}
    \includegraphics[width=\textwidth]{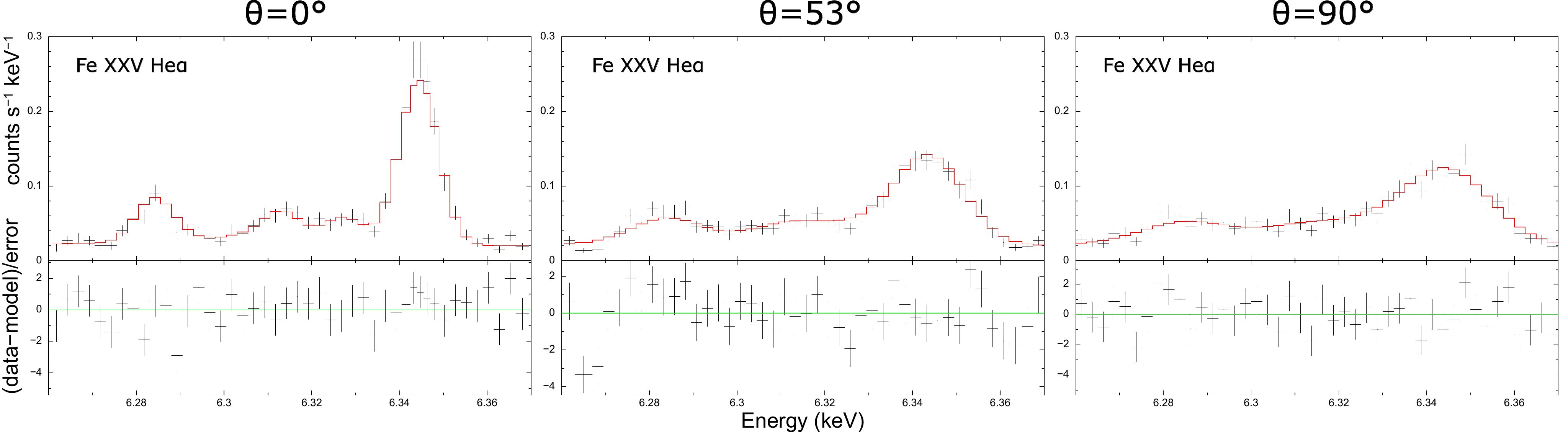}
  \end{center}
  \caption{
    XRISM mock spectra for the shocked ICM alone. Panels show results for viewing angles $\theta = 0^\circ$ (left), $53^\circ$ (middle), and $90^\circ$ (right). The upper part of each spectrum represents the normalized counts and best-fit \texttt{bapec} model curve, while the lower one represents the residuals divided by errors.
    {Alt text: Tree line graphs. Horizontal axis shows the energy from 6.26 to 6.37 kilo electron volt.}
    }\label{fig:fig3}
\end{figure*}

In this section, we analyze mock spectra of the shocked ICM region alone to investigate how the velocity dispersion depends on the viewing angle without the influence of the surrounding unshocked ICM. Spectral fitting is performed over the XRISM energy band of $2$--$10~\mathrm{keV}$. We use the C-statistic \citep{Cash1979} for model fitting, and quoted uncertainties represent $1\sigma$ confidence levels.
Figure~\ref{fig:fig3} presents the XRISM mock spectra for three different viewing angles focusing on He-like Fe. The curves show the best-fit \texttt{bapec} models. Table~\ref{tab:table2} compares the best-fit parameters with the directly estimated values (section~\ref{sec:direct}). The best-fit velocity dispersion is represented by $\sigma_{\mathrm{bulk,fit}}$.
The fitting results show that the estimated values of temperature, abundance and norm agree with the directly estimated values within $1\sigma$ uncertainties.

The best-fit bulk velocity dispersion ($\sigma_{\mathrm{bulk,fit}}$) increases from 136 to $442\rm\: km\: s^{-1}$ with viewing angle ($\theta$). Thus, the velocity dispersion is greater when the cocoon is observed from the side than when it is observed along the jet's direction, even though the cocoon's expansion speed is greater along the jet's direction. 
This indicates the high-speed region around the jet tip, which extends rapidly along the jet axis, contributes little to the X-ray spectra due to its low density and high temperature (figure~\ref{fig:fig1}). That is, at low densities, gas emissivity is low. At high temperatures, Fe ions are nearly fully ionized, and the emission lines are suppressed. 

\begin{table*}
  \caption{Best-fit parameters and directly estimated values for the shocked ICM alone.}
  \centering
  \begin{tabular}{lccc}
    \hline
    Viewing angle ($\theta$) & $0^\circ$ & $53^\circ$ & $90^\circ$ \\
    \hline
    \multicolumn{4}{l}{\textbf{Best-fit parameters}} \\
    $kT~(\mathrm{keV})$ & 
      $8.16^{+0.12}_{-0.12}$ & 
      $8.18^{+0.12}_{-0.12}$ &  
      $8.07^{+0.12}_{-0.12}$ \\
    Abundance ($Z_\odot$) & 
      $0.49^{+0.02}_{-0.01}$ &  
      $0.48^{+0.02}_{-0.02}$ & 
      $0.49^{+0.02}_{-0.02}$ \\
    $\bar{v}~(\mathrm{km~s^{-1}})$ \footnotemark[$*$]& 
      $6^{+8}_{-6}$ &  
      $35^{+16}_{-16}$ & 
      $-12^{+22}_{-20}$ \\
    $\sigma_{\mathrm{bulk,fit}}~(\mathrm{km~s^{-1}})$ & 
      $136^{+9}_{-9}$ &  
      $341^{+11}_{-11}$ & 
      $442^{+20}_{-18}$ \\
    norm ($\times 10^{-2}~\mathrm{cm^{-5}}$) & 
      $1.12^{+0.01}_{-0.01}$ & 
      $1.11^{+0.01}_{-0.01}$ &  
      $1.11^{+0.01}_{-0.01}$ \\
    C-stat / d.o.f & 
      12588 / 13620 & 
      12574 / 13676 & 
      12852 / 13573 \\
    \hline
    \multicolumn{4}{l}{\textbf{Directly estimated values}} \\
    $kT_{\mathrm{spec}}~(\mathrm{keV})$ & 8.04 & 8.04 & 8.04 \\
    $\sigma_{\mathrm{bulk,ew}}~(\mathrm{km~s^{-1}})$ & 367 & 460 & 505 \\
    $\sigma_{\mathrm{bulk,Fe}}~(\mathrm{km~s^{-1}})$ & 200 & 363 & 429 \\
    norm ($\times 10^{-2}~\mathrm{cm^{-5}}$) & 1.12 & 1.12 & 1.12 \\
    \hline
  \end{tabular}\label{tab:table2}
  \begin{tabnote}
    \parbox{\linewidth}{%
    \footnotemark[$*$]Best-fit velocity relative to Cyg A.}
  \end{tabnote}
\end{table*}

\subsection{Total ICM Mock Spectra}

\begin{figure*}
  \begin{center}
    \includegraphics[width=\textwidth]{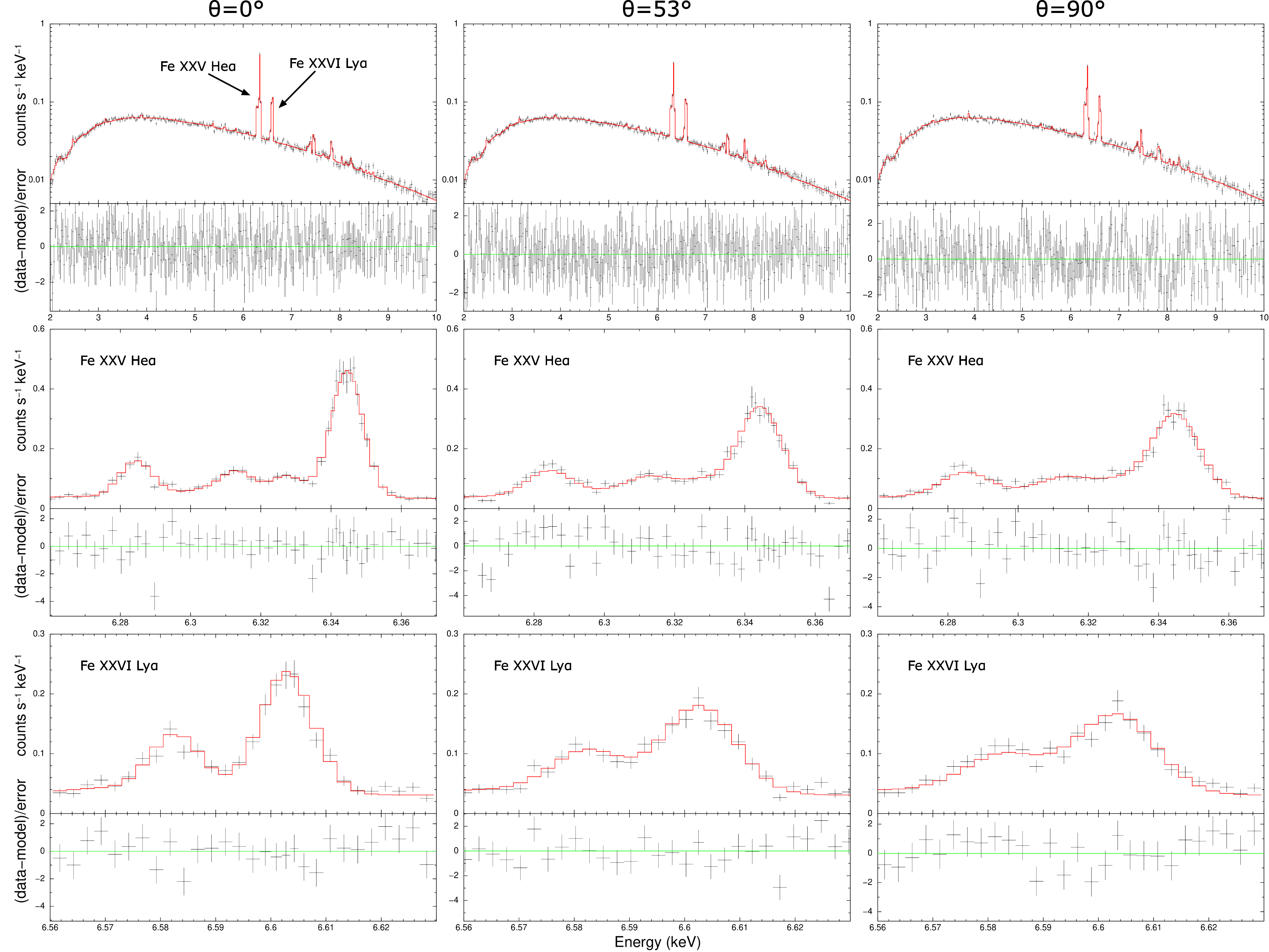}
  \end{center}
  \caption{
    XRISM mock spectra including both shocked and unshocked ICM. From left to right, the panels correspond to viewing angles $\theta = 0^\circ$, $53^\circ$, and $90^\circ$. The top column shows the full band ($2-10~\mathrm{keV}$), the center focuses on He-like Fe emission ($6.26-6.37~\mathrm{keV}$), and the bottom focuses on H-like Fe emission ($6.56-6.63~\mathrm{keV}$). The upper part of each spectrum represents the normalized counts and best-fit \texttt{bapec} model curve, while the lower one represents the residuals divided by errors. The emission lines broaden as the viewing angle increases, indicating larger velocity dispersions.
    {Alt text: Nine line graphs.}
  }\label{fig:fig4}
\end{figure*}

We now analyze the XRISM mock spectra including contributions from both the shocked and unshocked ICM. The fitting is performed over the $2-10~\mathrm{keV}$ range using the \texttt{bapec} model. 
Figure~\ref{fig:fig4} presents the total ICM mock spectra for three different viewing angles. The curves show the best-fit \texttt{bapec} models. Each column corresponds to a different viewing angle, and each row highlights different energy ranges including the full band and Fe line features. 

Table~\ref{tab:table3} presents the directly estimated values and best-fit parameters obtained from the total ICM mock spectra. $\sigma_{\mathrm{total,fit}}$ represent best fit total ICM velocity dispersion.

We now compare the total ICM results with those of the shocked ICM alone. The normalization in the total ICM case is approximately twice that of the shocked ICM, indicating that the X-ray contributions from the shocked and unshocked components are roughly comparable. This suggests that both components must be taken into account in actual observations.
The best-fit temperature for the total ICM is $\sim 7.6~\mathrm{keV}$, slightly lower than the $\sim 8.2~\mathrm{keV}$ obtained for the shocked ICM alone. This difference is attributed to the contribution from the lower-temperature unshocked ICM near the cluster center, which is assumed to be at $6~\mathrm{keV}$.

At $\theta = 0^\circ$, $\sigma_{\mathrm{total,fit}}$ increases to $\sim 151~\mathrm{km~s^{-1}}$, where the contributions from unshocked turbulence and shocked bulk motion become comparable. At $\theta = 53^\circ$ and $90^\circ$, $\sigma_{\mathrm{total,fit}}$ reaches $\sim 254~\mathrm{km~s^{-1}}$ and $\sim 287~\mathrm{km~s^{-1}}$, respectively, exceeding the values obtained from the shocked ICM alone.

As shown in Figures~\ref{fig:fig3} and \ref{fig:fig4}, the inclusion of the unshocked ICM reduces the relative variation in velocity dispersion with viewing angle. This result highlights the importance of considering both components when interpreting high-resolution X-ray spectra.

\subsection{Contribution of the Shocked ICM Tip Region}

The high-velocity regions around the jet terminals within the shocked ICM are rapidly heated and may emit line radiation under non-equilibrium ionization (NEI) conditions \citep{Prokhorov2010}. 
Our analysis suggests that the NEI contribution is expected to be minor, as the high-velocity components are characterized by low densities. For example, the flux (2--$10~\mathrm{keV}$) in the region with velocities above $1000~\mathrm{km~s^{-1}}$ (roughly corresponding to the region z > 50 kpc in figure \ref{fig:fig1}) accounts for only $13\%$ of the total cocoon flux. Furthermore, the averaged ionization timescale in this region is $\sim 5 \times 10^{12}~\mathrm{s~cm^{-3}}$, which is larger than the ionization timescale for heavy elements $\sim 10^{12}~\mathrm{s~cm^{-3}}$ \citep{Smith2010}. Therefore, assuming collisional ionization equilibrium, no significant difference is expected.

The 7--10 keV energy band is particularly sensitive to emission from hot plasma, as line radiation from gas with temperatures above 10 keV becomes relatively prominent in this range. Consequently, this band is expected to best capture the contribution of the high-temperature, high-velocity gas predicted by the simulations. We therefore performed a detailed spectral analysis of this interval, focusing in particular on the $\theta = 0^\circ$ configuration, where the line-of-sight component of the velocity is maximized. With future long-exposure time observations in mind, we adopted an exposure time of 1.5 Ms and fitted the spectra with both a single-component \texttt{bapec} model and a two-component \texttt{bapec+bapec} model. In the single-component fit, the obtained temperature was $kT = 7.61^{+0.16}_{-0.15}~\mathrm{keV}$, average velocity was $\bar{v} = 4^{+7}_{-10}~\mathrm{km\ s^{-1}}$ and the velocity dispersion was $\sigma_{\mathrm{bulk,fit}} = 161^{+11}_{-12}~\mathrm{km~s^{-1}}$. In contrast, in the two-component model, the low-temperature component was $kT_1 = 5.44^{+0.36}_{-1.06}~\mathrm{keV}$ and the high-temperature component was $kT_2 = 11.21^{+1.84}_{-0.71}~\mathrm{keV}$, with corresponding average velocities $\bar{v}_1 = -1^{+23}_{-28}~\mathrm{km\ s^{-1}}$ and $\bar{v}_2 = 7^{+31}_{-38}~\mathrm{km\ s^{-1}}$, respectively, and corresponding velocity dispersions $\sigma_{\mathrm{bulk,fit,1}} = 169^{+41}_{-37}~\mathrm{km~s^{-1}}$ and $\sigma_{\mathrm{bulk,fit,2}} = 145^{+56}_{-40}~\mathrm{km~s^{-1}}$, respectively.

To compare the goodness of fit between the two models, we rebinned the spectra to ensure a minimum of 25 counts per bin and evaluated the $\chi^2$ using the best-fit parameters obtained from the C-statistic optimization \citep{XRISM-Centaurus2025}. The resulting F-test gave a $p$-value of 0.43, indicating that the additional parameters in the two-component model do not provide a statistically significant improvement. Therefore, even when increasing the exposure time to capture the weak high-temperature emission lines, detecting a markedly larger velocity dispersion associated with a high-velocity component appears to be difficult.

\begin{table*}
  \caption{Directly estimated values and best-fit parameters for XRISM mock spectra including both shocked and unshocked ICM.}
  \label{tab:table3}
  \centering
  \begin{tabular}{lccc}
    \hline
    Viewing angle ($\theta$) & $0^\circ$ & $53^\circ$ & $90^\circ$ \\
    \hline
    \multicolumn{4}{l}{\textbf{Fitting parameters}} \\
    $kT~(\mathrm{keV})$ & 
      $7.63^{+0.09}_{-0.08}$ & 
      $7.66^{+0.09}_{-0.08}$ & 
      $7.58^{+0.08}_{-0.08}$ \\
    Abundance ($Z_\odot$) & 
      $0.50^{+0.01}_{-0.01}$ & 
      $0.50^{+0.01}_{-0.01}$ & 
      $0.49^{+0.01}_{-0.01}$ \\
    $\bar{v}~(\mathrm{km~s^{-1}})$ & 
      $-2^{+7}_{-4}$ &  
      $9^{+8}_{-9}$ & 
      $-16^{+10}_{-9}$ \\
    $\sigma_{\mathrm{total,fit}}~(\mathrm{km~s^{-1}})$ & 
      $151^{+6}_{-6}$ & 
      $254^{+8}_{-8}$ & 
      $287^{+10}_{-9}$ \\
    norm ($\times 10^{-2}~\mathrm{cm^{-5}}$) & 
      $2.01^{+0.01}_{-0.01}$ & 
      $2.00^{+0.01}_{-0.01}$ & 
      $2.01^{+0.01}_{-0.01}$ \\
    C-stat / d.o.f & 
      14199 / 14867& 
      14338 / 14914 & 
      14764 / 14804 \\
    \hline
    \multicolumn{4}{l}{\textbf{Directly estimated values}} \\
    $kT_{\mathrm{spec}}~(\mathrm{keV})$ & 7.59 & 7.59 & 7.59\\
    $\sigma_{\mathrm{total,ew}}~(\mathrm{km~s^{-1}})$ & 315 & 389 & 425 \\
    $\sigma_{\mathrm{total,Fe}}~(\mathrm{km~s^{-1}})$ & 170 & 224 & 250 \\
    norm ($\times 10^{-2}~\mathrm{cm^{-5}}$) & 2.01 & 2.01 & 2.01 \\
    \hline
  \end{tabular}
\end{table*}

\section{Summary}

In this study, we investigated the interaction between powerful AGN jets and the ICM using the Cyg A as a model, through two-dimensional hydrodynamic simulations and a mock observation using XRISM/Resolve. The simulations reproduce characteristic jet-induced structures such as forward shocks, hot cocoons, and compressed dense gas, which together shape the dynamical and thermal state of the ICM.

Mock spectra were constructed by dividing the simulated ICM into shocked and unshocked components and applying the instrumental response of XRISM/Resolve. We performed the mock observations from three different viewing angles, $\theta = 0^\circ, 53^\circ,$ and $90^\circ$, to evaluate the impact of projection effects on the observable line widths. The results show a clear dependence of the measured velocity dispersion on the viewing angle. When observed along the jet axis ($\theta=0^\circ$), the velocity dispersion appears significantly smaller than expected, because the most rapidly expanding jet-tip region is both hot and tenuous, and thus contributes little to the thermal bremsstrahlung and line emission  radiation. For larger viewing angles, the contribution of the shocked ICM to the emission becomes more prominent, leading to systematically larger velocity dispersions.

These results indicate that the detectability of bulk motions and turbulence with XRISM depends not only on the intrinsic jet--ICM interaction but also strongly on the viewing geometry and the thermodynamic structure of the emitting gas. In particular, the reduced velocity dispersion seen along the jet axis highlights the importance of considering both thermal and geometrical effects when interpreting high-resolution X-ray spectra of radio galaxies such as Cyg A. This study provides theoretical guidance for upcoming XRISM observations and contributes to a deeper understanding of AGN feedback in clusters.  

\section*{Acknowledgments}

We thank Prof. Kazuhiro Nakazawa for his insightful suggestions on the possible contribution of NEI plasma, which helped improve the clarity of this manuscript.

\vspace{5mm}

\noindent
\textit{Note added in proof.}
After the submission of this paper, XRISM observations of the Cyg~A have been reported \citep{Majumder2025}.
A LOS velocity dispersion of
$\sigma_v = 261^{+13}_{-13}~\mathrm{km\,s^{-1}}$
was measured in the central region using XRISM/Resolve.
In our mock XRISM observation assuming the same field of view
($6 \times 6$ pixels), we obtained a slightly smaller velocity dispersion, $\sigma_{\rm total,fit} = 238^{+7}_{-7}~\mathrm{km\,s^{-1}}$. This difference is likely attributable to the use of two-dimensional numerical simulations in this work, which tend to underestimate turbulent gas motions compared to fully three-dimensional simulations. Nevertheless, the mock observation reproduces the observed velocity dispersion and captures the essential characteristics of AGN-driven gas motions in Cyg~A.


\end{document}